\documentclass[iop]{emulateapj}

\usepackage{natbib}

\newcommand{\kms}{km\,s$^{-1}$}

\newcommand{\msun}{$M_\sun$}

\newcommand{\gp}{$g_\mathrm{P1}$}
\newcommand{\rp}{$r_\mathrm{P1}$}
\newcommand{\ip}{$i_\mathrm{P1}$}
\newcommand{\zp}{$z_\mathrm{P1}$}
\newcommand{\yp}{$y_\mathrm{P1}$}
\newcommand{\mbh}{$M_\mathrm{BH}$}
\newcommand{\mgii}{\ion{Mg}{2}}

\slugcomment{Accepted for publication in ApJ Letters}

\shorttitle{Three quasars at $6.5<z<6.7$ in PS1}
\shortauthors{Venemans et al.}

\begin{document}

\title{THE IDENTIFICATION OF $z$-DROPOUTS IN PAN-STARRS1: THREE QUASARS AT
  $6.5<z<6.7$\altaffilmark{1}
}

\author{B. P. Venemans\altaffilmark{2}, 
E. Ba{\~n}ados\altaffilmark{2},
R. Decarli\altaffilmark{2}, 
E.~P. Farina\altaffilmark{2},
F. Walter\altaffilmark{2}, 
K. C. Chambers\altaffilmark{3}, 
X. Fan\altaffilmark{4}, 
H-W. Rix\altaffilmark{2},
E. Schlafly\altaffilmark{2}, 
R. G. McMahon\altaffilmark{5,6}, 
R. Simcoe\altaffilmark{7},
D. Stern\altaffilmark{8}, 
W. S. Burgett\altaffilmark{9}, 
P. W. Draper\altaffilmark{10},
H. Flewelling\altaffilmark{3}, 
K. W. Hodapp\altaffilmark{3},
N. Kaiser\altaffilmark{3}, 
E. A. Magnier\altaffilmark{3}, 
N. Metcalfe\altaffilmark{10},
J. S. Morgan\altaffilmark{3},
P. A. Price\altaffilmark{11},
J. L. Tonry\altaffilmark{3},
C. Waters\altaffilmark{3},
Y. AlSayyad\altaffilmark{12},
M. Banerji\altaffilmark{5,6},
S. S. Chen\altaffilmark{7},
E. A. Gonz\'alez-Solares\altaffilmark{5},
J. Greiner\altaffilmark{13},
C. Mazzucchelli\altaffilmark{2},
I. McGreer\altaffilmark{4}, 
D. R. Miller\altaffilmark{7},
S. Reed\altaffilmark{5},
P. W. Sullivan\altaffilmark{7}
}

\altaffiltext{1}{Based in part on observations collected at the European
  Southern Observatory, Chile, programs 179.A-2010, 092.A-0150, 093.A-0863 and
  093.A-0574, and at the Centro Astron{\'o}mico Hispano Alem{\'a}n (CAHA) at
  Calar Alto, operated jointly by the Max-Planck Institut f{\"u}r Astronomie
  and the Instituto de Astrof{\'i}sica de Andaluc{\'i}a (CSIC). This paper
  also includes data gathered with the 6.5 meter Magellan Telescopes located
  at Las Campanas Observatory, Chile, at the MMT Observatory, a joint facility
  of the University of Arizona and the Smithsonian Institution, and with the
  LBT. }  
\altaffiltext{2}{Max-Planck Institute for Astronomy, K{\"o}nigstuhl 17, 69117
  Heidelberg, Germany}
\altaffiltext{3}{Institute for Astronomy, University of Hawaii, 2680 Woodlawn
  Drive, Honolulu, HI 96822, USA }
\altaffiltext{4}{Steward Observatory, The University of Arizona, 933 North
  Cherry Avenue, Tucson, AZ 85721-0065, USA}
\altaffiltext{5}{Institute of Astronomy, University of Cambridge, Madingley
  Road, Cambridge, CB3 0HA, UK}
\altaffiltext{6}{Kavli Institute for Cosmology, University of Cambridge,
  Madingley Road, Cambridge CB3 0HA, UK}
\altaffiltext{7}{MIT-Kavli Center for Astrophysics and Space Research, 77
  Massachusetts Avenue, Cambridge, MA 02139, USA}
\altaffiltext{8}{Jet Propulsion Laboratory, California Institute of
  Technology, 4800 Oak Grove Drive, Mail Stop 169-221, Pasadena, CA 91109,
  USA}
\altaffiltext{9}{GMTO Corporation, 251 S. Lake Ave., Suite 300, Pasadena, CA
  91101, USA}
\altaffiltext{10}{Department of Physics, Durham University, South Road, Durham
  DH1 3LE, UK}
\altaffiltext{11}{Department of Astrophysical Sciences, Princeton University,
  Princeton, NJ 08544, USA}
\altaffiltext{12}{Department of Astronomy, University of Washington, Box
  351580, Seattle, WA 98195, USA}
\altaffiltext{13}{Max-Planck-Institut f\"ur extraterrestrische Physik,
  Giessenbachstrasse 1, 85748 Garching, Germany}

\begin{abstract}
  Luminous distant quasars are unique probes of the high redshift
  intergalactic medium (IGM) and of the growth of massive galaxies and
  black holes in the early universe. Absorption due to neutral
  Hydrogen in the IGM makes quasars beyond a redshift of $z\simeq6.5$
  very faint in the optical $z$-band, thus locating quasars at higher
  redshifts require large surveys that are sensitive above 1
  micron. We report the discovery of three new $z>6.5$ quasars,
  corresponding to an age of the universe of $<850$\,Myr, selected as
  $z$-band dropouts in the Pan-STARRS1 survey. This increases the
  number of known $z>6.5$ quasars from 4 to 7. The quasars have
  redshifts of $z=6.50$, 6.52, and 6.66, and include the brightest
  $z$-dropout quasar reported to date, PSO J036.5078+03.0498 with
  $M_{1450}=-27.4$. We obtained near-infrared spectroscopy for the
  quasars and from the \ion{Mg}{2} line we estimate that the central
  black holes have masses between $5\times10^8$ and
  $4\times10^9$\,\msun, and are accreting close to the Eddington limit
  ($L_\mathrm{Bol}/L_\mathrm{Edd}=0.13-1.2$). We investigate the
  ionized regions around the quasars and find near zone radii of
  $R_\mathrm{NZ}=1.5-5.2$\,proper Mpc, confirming the trend of decreasing
  near zone sizes with increasing redshift found for quasars at
  $5.7<z<6.4$. By combining $R_\mathrm{NZ}$ of the PS1 quasars with
  those of $5.7<z<7.1$ quasars in the literature, we derive a
  luminosity corrected redshift evolution of
  $R_\mathrm{NZ,corrected}=(7.2\pm0.2)-(6.1\pm0.7)\times(z-6)$\,Mpc. However,
  the large spread in $R_\mathrm{NZ}$ in the new quasars implies a
  wide range in quasar ages and/or a large variation in the neutral
  Hydrogen fraction along different lines of sight.
\end{abstract}


\section{INTRODUCTION}

Quasars are the most luminous non-transient objects known. Their high
luminosity makes quasars ideal to probe the universe at early cosmic
times. Since distant ($z\gtrsim5.7$) luminous quasars are rare, with an
estimated source density of $\sim$1 per Gpc$^3$ \citep[e.g.,][]{fan04,wil10a},
surveys covering a large area of the sky are required to uncover the distant
quasar population. Over the last 15 years more than 70 quasars with redshifts
between $5.5<z<6.5$ have been discovered in various surveys
\citep[e.g.,][]{fan06b,jia08,mor09,wil10b,mor12,ban14}. Most of these quasars
have been found by looking for sources with a large break between the optical
$i$ and $z$-bands \citep[e.g.,][]{fan06b}, the so-called $i$-band dropouts or
$i$-dropouts. To find quasars beyond $z\sim6.5$ wide-field surveys with
coverage beyond $\sim$1\,$\mu$m are needed.

Currently, four quasars above $z>6.5$ have been discovered in
near-infrared surveys. \citet{mor11} presented a quasar at $z=7.1$
discovered in the UK infrared Telescope Infrared Deep Sky Survey
\citep[UKIDSS;][]{law07}, while \citet{ven13} reported three quasars at
$6.6<z<6.9$ from the Visible and Infrared Survey Telescope Kilo-Degree
Infrared Galaxy (VIKING) survey. Detailed studies of these four
$z>6.5$ quasars have given insight into the properties of the universe
less than a Gyr after the Big Bang. For example, the optical spectrum
of the $z=7.1$ quasar places constraints on the fraction of neutral
Hydrogen \citep{mor11,bol11}, while \citet{sim12} uses near-infrared
spectroscopy to put limits on the metal enrichment (``metallicity'')
of the intergalactic medium up to $z\sim7$. Furthermore, these quasars
set a lower limit to the number density of supermassive black holes at
$z>6.5$: $\rho(M_\mathrm{BH}>10^9\,M_\odot) >
1.1\times10^{-9}$\,Mpc$^{-3}$ \citep{ven13,der14}.

The limitation of both UKIDSS and VIKING is that they cover only small
fractions of the extragalactic sky: UKIDSS has imaged $\sim$4000\,deg$^2$,
while VIKING will cover an area of 1500\,deg$^2$. Overcoming this limitation,
the Panoramic Survey Telescope \& Rapid Response System 1 \citep[Pan-STARRS1,
PS1;][]{kai02,kai10} has imaged the whole sky above a declination of
--30$^\circ$ for about four years in five filters \citep[\gp, \rp, \ip, \zp,
and \yp;][]{stu10,ton12}. The inclusion of the \yp\ filter \citep[central
wavelength $\lambda_c=9620$\,\AA, FWHM\,$=890$\,\AA;][]{ton12} enables the
search for luminous quasars at $z>6.5$ over more than 20,000\,deg$^2$ of
extragalactic sky by selecting sources with a red \zp--\yp\ colors
($z$-dropouts, see Fig.\ \ref{spectra}). In this paper we present the first
results from our ongoing search for $z>6.5$ quasars in the PS1 data.

All magnitudes are given in the AB system and we adopt a cosmology with
$H_0=70$ km\,s$^{-1}$\,Mpc$^{-1}$, $\Omega_M=0.28$ and $\Omega_\lambda=0.72$
\citep{kom11}.

\section{CANDIDATE SELECTION}
\label{canselection}

\subsection{The Pan-STARRS1 3$\pi$ Survey}

To search for $z$-dropouts, we made use of the first internal release
of stacked PS1 imaging data (PV1). We selected our initial $z>6.5$
quasar candidates as objects with a signal-to-noise ratio
(S/N)$_{y\mathrm{P1}} > 7$, and a non-detection in the \gp\, \rp\, and
\ip\ bands (i.e.\ S/N\,$<3$). We further required a break between the
\zp\ and \yp\ bands \citep[see also Fig.\ 1 in][]{ban14}:

\begin{displaymath}
  \mathrm{(S/N)}_{z\mathrm{P1}}>3~\mathrm{AND}~(z_\mathrm{P1}-y_\mathrm{P1}>
  1.4)~~~~\mathrm{OR}
\end{displaymath}
\begin{displaymath}
  \mathrm{(S/N)}_{z\mathrm{P1}}<3~\mathrm{AND}~(z_\mathrm{P1,lim,3\sigma}-y_\mathrm{P1}> 1.4).
\end{displaymath}

\noindent
To avoid extended sources we demanded that the difference between the
\yp-band PSF and aperture magnitudes ($y_\mathrm{ext}$) to be
consistent within 0.3\,mag. This value was determined by comparing
$y_\mathrm{ext}$ with spectroscopically confirmed stars and galaxies
from the SDSS-III database\footnote{http://www.sdss3.org/}. Setting
$y_\mathrm{ext}<0.3$ selected the vast majority of stars ($>85$\%),
while rejecting a large fraction ($>$94\%) of galaxies. Lastly, we
removed sources that were marked as likely spurious in the catalogs
\citep[see][]{ban14} or that had less than 85\% of the expected flux
in the \zp\ or \yp\ images on valid pixels. For bright sources
(\yp\,$< 19.5$) we applied the same criteria but we relaxed our limits
in the \gp, \rp, and \ip-bands, requiring S/N\,$<5$.

The total number of $z$-dropouts selected from the PS1 catalogs after
removing objects in the plane of the Milky Way and M31 ($\left| b \right| <
20^\circ$ and $7^\circ< \,$R.A.$ <14^\circ$; $37^\circ< \,$Decl.$ <43^\circ$)
was $328,372$ (of which 13,093 had \yp\,$< 19.5$).

\subsection{Public infrared surveys}
\label{surveys}

To extend and verify the photometry of the quasar candidates selected from the
PS1 catalogs we first matched the sources with several public infrared surveys.

{\it UKIDSS:} The PS1 candidates were matched with the near-infrared data of
the UKIDSS survey \citep{law07}. The UKIDSS Large Area Survey (LAS) provides
$Y$, $J$, $H$ and $K$ imaging over $\sim$4000 deg$^2$. We matched the PS1
$z$-dropout list with the catalogs from the UKIDSS data release\footnote{
  \url{http://surveys.roe.ac.uk/wsa/dr10plus\_release.html}} 10, using a
search radius of 2\farcs0. We identified objects as foreground interlopers
if they had a $Y-J>0.6$ or $y_\mathrm{P1}-J>1$ \citep[which is typical for
cool dwarfs; see e.g.,][]{bes13} and removed them from our candidate lists.

{\it VHS:} For objects in the area $150^\circ<$\,R.A.\,$<240^\circ$
and $-20^\circ<$\,Decl.\,$<0^\circ$, aperture photometry was performed
on the $Y$, $J$, $H$ and $K_s$ images of the VISTA Hemisphere Survey
\citep[VHS;][]{mcm13}. We applied the same color criteria as for our
UKIDSS matched list.

{\it WISE:} The {\em Wide-field Infrared Survey Explorer} \citep[{\em
  WISE};][]{wri10} surveyed the entire mid-infrared sky in four bands
centered at 3.4, 4.6, 12 and 22 $\mu$m. The NEOWISE observations
\citep{mai11} surveyed 70\% of the sky at 3.4 and 4.6 $\mu$m
(hereafter W1 and W2). Both surveys were combined to produce the
AllWISE catalogs\footnote{
  \url{http://wise2.ipac.caltech.edu/docs/release/allwise/}}. To rule
out spurious candidates, we required PS1 $z$-dropouts without a match
in the VHS or UKIDSS surveys to have a counterpart in the AllWISE
catalogs within 3\farcs0 to be considered real sources.

Objects with a S/N\,$>3$ in W1 and W2 were assigned a higher priority
if their colors fulfilled the additional criteria:
$-0.2<W_{1,\mathrm{AB}}-W_{2,\mathrm{AB}}<0.86$ AND
$W_{1,\mathrm{AB}}-W_{2,\mathrm{AB}} >
-1.45\times(y_\mathrm{P1}-W_{1,\mathrm{AB}})-0.455$, which is loosely
based on the PS1-{\it WISE} colors of brown dwarfs
\citep[e.g.,][]{bes13}. Objects with a S/N\,$<3$ in W1 or W2 were
assigned with an intermediate priority and the remaining candidates
were given a low priority.

For the approximately 13,000 objects with a match in at least one of
the above surveys, we performed forced photometry on the PS1
images to confirm the colors and non-detections
\citep[see][]{ban14}. After visually inspecting the remaining
$\sim$1000 candidates we selected the best $\sim$500 objects that were
our main targets for follow-up observations.

\section{FOLLOW-UP OBSERVATIONS}
\label{follow-up}

\begin{deluxetable*}{lccccc}
  \tabletypesize{\scriptsize}
  \tablecaption{Imaging and Spectroscopic Observations of Quasar
    Candidates \label{specobs}} \tablewidth{0pt}
  \tablehead{\colhead{Object\tablenotemark{a}} & \colhead{Date} &
    \colhead{Telescope/Instrument} & \colhead{$\lambda$ range / filters} &
    \colhead{Exposure Time} & \colhead{Slit Width} }
  \startdata 
  {} & 2014 January 24--February 5 & MPG 2.2m/GROND & $g^\prime$, $r^\prime$, 
  $i^\prime$, $z^\prime$, $J$, $H$, $K$ & 460\,s--960\,s & -- \\
  {} & 2014 March 2--6 & NTT/EFOSC2 & $I_N$, $Z_N$ & 600\,s & -- \\ 
  {} & 2014 March 2 \& 5 & NTT/SofI & $J$ & 300\,s & -- \\ 
  {} & 2014 March 16--19 & CAHA 3.5m/Omega2000 & $z$, $Y$, $J$ & 
  300\,s--600\,s & -- \\ 
  {} & 2014 July 23--28 & NTT/EFOSC2 & $I_N$, $Z_N$ & 600\,s & -- \\
  {} & 2014 July 25 & NTT/SofI & $J$ & 600\,s & -- \\
  {} & 2014 August 7 \& 11--13 & CAHA 3.5m/Omega2000 & $Y$, $J$ & 600\,s 
  & -- \\
  \hline
  P167--13 & 2014 April 26 & VLT/FORS2 & 0.74--1.07\,$\mu$m
  & 2630\,s & 1\farcs3 \\
  {} & 2014 May 30--June 2 & Magellan/FIRE & 0.82--2.49\,$\mu$m &
  12004\,s & 0\farcs6 \\
  P036+03 & 2014 July 25 & NTT/EFOSC2 & 0.60--1.03\,$\mu$m & 7200\,s
  & 1\farcs2 \\
  {} & 2014 September 4--6 & Magellan/FIRE & 0.82--2.49\,$\mu$m &
  8433\,s & 0\farcs6 \\
  {} & 2014 October 20 & Keck I/LRIS & 0.55--1.03\,$\mu$m & 
  1800\,s\tablenotemark{b} & 1\farcs0 \\
  P338+29 & 2014 October 19 & MMT/Red Channel & 0.67--1.03\,$\mu$m &
  1800\,s & 1\farcs0 \\
  {} & 2014 October 30 & Magellan/FIRE & 0.82--2.49\,$\mu$m & 
  7200\,s\tablenotemark{b} & 0\farcs6 \\
  {} & 2014 November 27 & LBT/MODS & 0.51--1.06 $\mu$m & 
  2700\,s & 1\farcs2 \\
  {} & 2014 December 6 & LBT/LUCI & 2.05--2.37 $\mu$m & 
  3360\,s & 1\farcs5
  \enddata
  \tablenotetext{a}{For the full name and coordinates, see Table
    \ref{qsocolors}.}
  \tablenotetext{b}{Observations in cloudy conditions.}
\end{deluxetable*}

To confirm the colors of the possible quasars and to remove lower
redshift interlopers, we imaged 194 $z$-dropout candidates during five
observing runs. We obtained optical and infrared images between 2014
January 24 and 2014 August 13 with the MPG 2.2m/GROND
\citep[][]{gre08}, NTT/EFOSC2 \citep{buz84}, NTT/SofI \citep{moo98}
and the Calar Alto 3.5m/Omega2000 \citep{bai00}, see Table
\ref{specobs} for the details of the observations.

Candidates were considered foreground interlopers if they had
$y_\mathrm{P1}-J>1$ (see Sect.\ \ref{surveys}). Sources with a
$y_\mathrm{P1}-J < -1$ or undetected in $J$ were rejected on the basis
that they could be moving, varying or a spurious object in the PS1
catalog. We reobserved objects with $-1.0<y_\mathrm{P1}-J<1.0$ with
the NTT in the filters $I_N$ and $Z_N$. Only three sources remained
undetected in $I_N$ or were red with $I_N-Z_N\gtrsim2$. These objects
were targets for spectroscopy.

\begin{figure*}
\includegraphics[width=\textwidth]{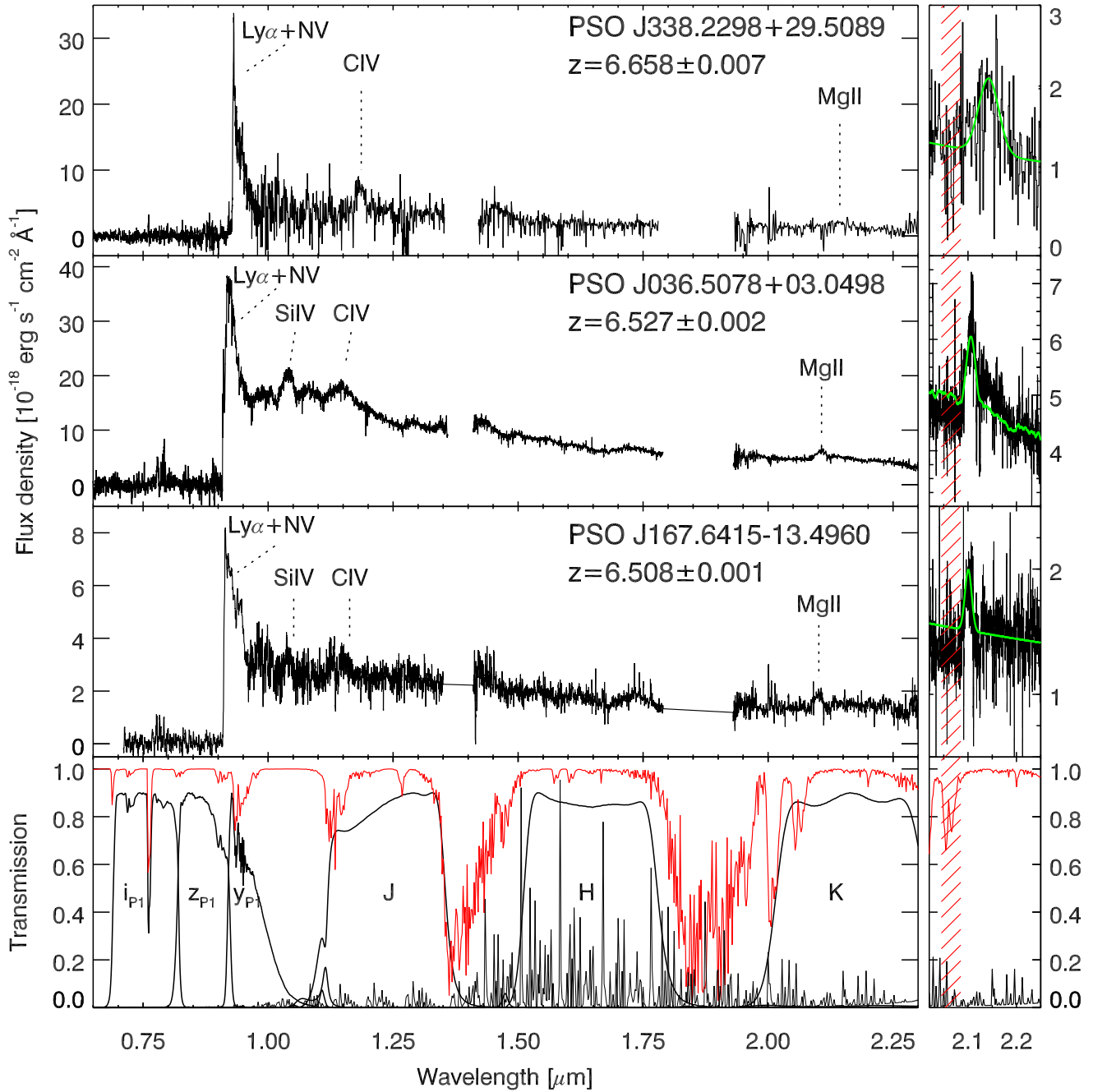}
\caption{Merged optical and near-infrared spectra of the three new
  $z>6.5$ quasars. The position of various strong, broad emission
  lines are indicated. The bottom panel shows the transmission curves
  of the \ip, \zp, and \yp\ filters, and the GROND $J$, $H$ and $K$
  filters. Also plotted is the relative strength of the sky emission
  (in black) and the telluric absorption spectrum (in red). The panels
  to the right show a zoom on the \ion{Mg}{2} line with the model best
  fitting the line and continuum overplotted in green. The red stripes
  mark a region with low sky transparency that was excluded from the
  fitting.}
\label{spectra}
\end{figure*}

We obtained optical and near-infrared spectroscopy of all the three
candidates that had good quasar colors after the follow-up
imaging. We carried out spectroscopic observations between 2014
April 26 and 2014 December 6 using the following instruments:
VLT/FORS2 \citep{app98}; Megallan/FIRE \citep{sim08,sim10};
NTT/EFOSC2; Keck/LRIS \citep{oke95}; MMT/Red Channel Spectrograph; and
LBT/MODS \citep{pog10} and LBT/LUCI \citep{sei03}. Details of the
observations are listed in Table \ref{specobs}.

We reduced the data following standard reduction steps
\citep[e.g.,][]{ven13,ban14}. We show the merged spectra in Fig.\
\ref{spectra}.

\section{THREE QUASARS AT $z>6.5$}
\label{3quasars}

\begin{deluxetable*}{lccc}
  \tabletypesize{\footnotesize}
  \tablecaption{Photometric Properties and Derived Parameters of the New
    $z>6.5$ Quasars From PS1. 
    \label{qsocolors}} 
  \tablewidth{0pt} 
  \tablehead{\colhead{} & \colhead{PSO J167.6415--13.4960} & \colhead{PSO
      J036.5078+03.0498} & \colhead{PSO J338.2298+29.5089}} \startdata
  Abbreviated name & P167--13 & P036+03 & P338+29 \\
  R.A.\ (J2000) & 11$^h$10$^m$33$^s\!\!$.98 & 02$^h$26$^m$01$^s\!\!$.88 &
  22$^h$32$^m$55$^s\!\!$.15 \\
  Decl.\ (J2000) & --13$^\circ$29\arcmin45\farcs6 &
  +03$^\circ$02\arcmin59\farcs4 & +29$^\circ$30\arcmin32\farcs2 \\
  $I_{N}$ & $>$24.68\tablenotemark{a} & $23.62\pm0.18$ &
  $>$24.37\tablenotemark{a} \\
  $z_\mathrm{P1}$ & $>$22.54\tablenotemark{a} & $21.51\pm0.24$ &
  $>$22.25\tablenotemark{a} \\
  $Z_{N}$         & $22.08\pm0.09$ & $20.46\pm0.04$ & $21.92\pm0.09$ \\
  $y_{\mathrm{P1}}$ & $20.49\pm0.12$ & $19.37\pm0.04$ & $20.13\pm0.08$ \\
  $J$         & $21.21\pm0.09$ & $19.51\pm0.03$ & $20.74\pm0.09$ \\
  $W_{1,\mathrm{AB}}$ & $21.13\pm0.39$\tablenotemark{b} & $19.43\pm0.08$ & 
  $20.51\pm0.21$ \\
  $M_\mathrm{1450}$ & $-25.58\pm0.13$ & $-27.36\pm0.03$ & $-26.04\pm0.09$ \\
  $\beta$ & $-1.0\pm0.1$ & $-1.70\pm0.05$ & $-1.85^{+0.08}_{-0.05}$ \\
  $z_\mathrm{MgII}$ & $6.508\pm0.001$ & $6.527\pm0.002$ &
  $6.658\pm0.007$ \\
  FWHM$_\mathrm{MgII}$ (\kms) & $2350\pm470$ & $3500^{+1010}_{-740}$ & 
  $6800^{+1200}_{-900}$ \\
  \mbh\ (\msun) & $(4.9\pm2.0)\times10^8$ & $(1.9^{+1.1}_{-0.8})\times10^9$ & 
  $(3.7^{+1.3}_{-1.0})\times10^9$ \\
  $L_\mathrm{Bol,3000\,\mathring{A}}/L_\mathrm{Edd}$ & $1.2\pm0.5$ & 
  $0.96_{-0.35}^{+0.70}$ & $0.13^{+0.05}_{-0.04}$ \\ 
  $R_\mathrm{NZ}$ (proper Mpc) & $1.5\pm0.7$ & $3.1\pm0.7$ & $5.2\pm0.7$ \\
  $R_\mathrm{NZ,corrected}$ (proper Mpc) & $2.3\pm1.1$ & $2.8\pm0.6$ & 
  $6.9\pm0.9$
\enddata
\tablenotetext{a}{Non-detections listed as $3\sigma$ upper limits.}
\tablenotetext{b}{\,$\!$Listed in the AllWISE Reject table.}
\end{deluxetable*}

All three $z$-dropouts for which we obtained optical spectroscopy
showed a strong continuum break in their optical spectrum (Fig.\
\ref{spectra}) and were identified as quasars at redshifts
$6.5<z<6.7$. We fitted the continuum with three components: a
power-law with slope $\beta$ ($f_\lambda \sim \lambda^\beta$), a
Balmer continuum and an \ion{Fe}{2} template \citep[see
e.g.,][]{der14}. In all cases the Balmer continuum was found to be
negligible at the wavelengths we considered ($\lambda_\mathrm{rest}
\lesssim 3000$\,\AA). A single-Gaussian fit of the emission lines
(most prominently \ion{C}{4} and \ion{Mg}{2}) provided a sufficiently
good model of the line profiles given the S/N of our spectra. Only in
the spectrum of the brightest quasar, PSO J036.5078+03.0498, we were
able to constrain on the \ion{Fe}{2} emission (Section
\ref{p036}). The near-infrared spectra of the other two quasars did
not have sufficient S/N. This did not significantly affect the fit of
the \ion{Mg}{2} lines in these quasars.

The redshifts were determined by the peak of the \ion{Mg}{2}
line. Other bright emission lines, such as \ion{Si}{4} $\lambda$ 1397
and \ion{C}{4} $\lambda$ 1549, are blueshifted by 300--2000
km\,s$^{-1}$ with respect to \ion{Mg}{2}. Such shifts are similar to
those measured in spectra of other distant luminous quasars
\citep[e.g.,][]{ric02b,der14}.

We estimated black hole masses using the local scaling relation
based on the \ion{Mg}{2} line \citep[Eq.\ 1 in][]{ves09}, which has a
systematic uncertainty of a factor $\sim$3. The black hole mass
uncertainties quoted in Sections \ref{p167}--\ref{p338} and in Table
\ref{qsocolors} represent only statistical errors.

We computed bolometric luminosities by applying the bolometric
correction obtainted by \citet{she08} to the monochromatic luminosity
density measured at 3000\,\AA. The Eddington luminosity is defined as
$L_\mathrm{Edd}=1.3\times10^{38}\,
(M_\mathrm{BH}/M_\sun)$\,erg\,s$^{-1}$.

A summary of the photometric properties and the parameters derived
from the spectra is provided in Table \ref{qsocolors}. Below we
describe the new quasars in more detail.

\subsection{PSO J167.6415--13.4960}
\label{p167}

The quasar PSO J167.6415--13.4960 (hereafter P167--13) was discovered based on
forced photometry on VHS images at the positions of PS1 candidates. Our FORS2
discovery spectrum revealed a source with a strong continuum decrement around
9100\,\AA, and we identified the object as a quasar with a redshift of
$z\sim6.52$. From the near-infrared spectrum we derive a redshift
$z_\mathrm{MgII}=6.508\pm0.001$. This quasar is the faintest of our new
discoveries with $M_\mathrm{1450}=-25.58\pm0.13$. The power-law slope of
$\beta=-1.0\pm0.1$ is red compared to the slope of $\beta=-1.3$ of the SDSS
quasar composite spectrum of \citet{van01}. The estimated black hole mass is
\mbh$_\mathrm{,MgII}\approx(4.9\pm 2.0)\times10^8$\,\msun, and the Eddington
ratio is consistent with maximal accretion
($L_\mathrm{Bol}/L_\mathrm{Edd}=1.2\pm0.5$).

\subsection{PSO J036.5078+03.0498}
\label{p036}

PSO J036.5078+03.0498 (hereafter P036+03) was selected as part of our
extended, bright $z$-dropout search and was matched to a source in the
UKIDSS and {\it WISE} catalogs. The high S/N FIRE spectrum revealed
blue quasar continuum emission ($\beta=-1.70\pm0.05$) at a redshift
$z_\mathrm{MgII}=6.527\pm0.002$. The absolute magnitude of
$M_\mathrm{1450}=-27.36\pm0.03$ makes this quasar one of the most
luminous objects known at $z>6$. The bolometric luminosity is
estimated to be
$L_\mathrm{Bol,3000\,\mathring{A}}=(2.38\pm0.09)\times10^{47}$\,erg\,s$^{-1}$. The
central black hole has an estimated mass of \mbh$_\mathrm{,MgII}=
(1.9^{+1.1}_{-0.8})\times10^9$\,\msun. The accretion rate is close to
Eddington with $L_\mathrm{Bol}/L_\mathrm{Edd} = 0.96\pm0.55$. The
quality of the infrared spectrum is not sufficient to constrain the
\ion{Fe}{2} emission to better than $2\sigma$. We measure
\ion{Fe}{2}/\ion{Mg}{2}\,$=3.4\pm1.7$, fully consistent with
previously discovered quasars at similar redshifts
\citep[e.g.,][]{der14}.

\subsection{PSO J338.2298+29.5089}
\label{p338}

PSO J338.2298+29.5089 (hereafter P338+29) was one of the $z$-dropout
candidates with a match in the {\it WISE} catalog. The discovery
spectrum shows a source with a strong, narrow emission line at
$\sim$9314\,\AA\ and continuum redward of the line, which we identify
as Ly$\alpha$ at a redshift of $z=6.66$. From the FIRE spectrum we
measure $z_\mathrm{MgII}=6.658\pm0.007$, $M_{1450}=-26.04\pm0.09$, and
a blue continuum slope $\beta=-1.85^{+0.08}_{-0.05}$. Although the
\ion{Mg}{2} line suffers from sky residuals on its blue side, we
estimate a black hole mass of
$M_\mathrm{BH,MgII}=(3.7^{+1.3}_{-1.0})\times10^9$\,\msun\ and an
accretion rate of
$L_\mathrm{Bol}/L_\mathrm{Edd}=0.13^{+0.05}_{-0.04}$.

\section{QUASAR IONIZATION REGION}
\label{nearzones}

\begin{figure*}
\includegraphics[width=\textwidth]{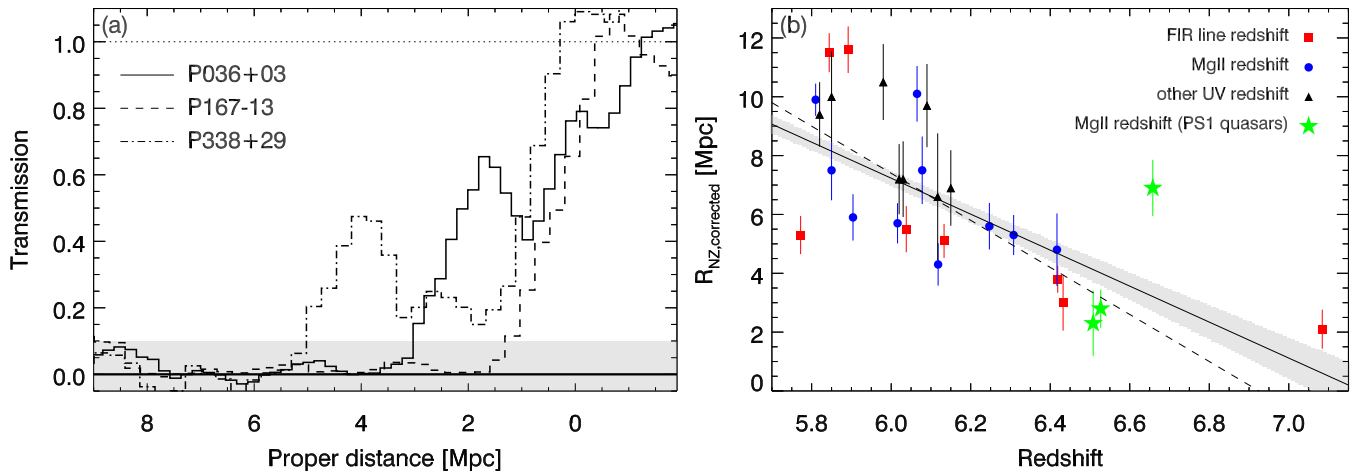}
\caption{(a) The near zone of the quasars in physical (proper)
  Mpc. The transmission was computed by normalizing the observed
  spectra with a power-law continuum and two Gaussians fitted to the
  Ly$\alpha$ and \ion{N}{5} lines. The resulting normalized spectra
  were subsequently smoothed to a resolution of 20\,\AA. The
  redshifted position of the Ly$\alpha$ line, based on the \mgii\
  redshift, is taken as zeropoint on the x-axis. The grey region
  indicates $<$10\% transmission. The ionized region around P167--13
  extends to only $\sim$1.5\,Mpc, while the near zones of P036+03 and
  P338+29 are 3.1\,Mpc and 5.2\,Mpc, respectively. (b) Luminosity
  corrected near zone radii as function of redshift for $z>5.7$
  quasars. Red squares, blue circles and black triangles represent
  measurements taken from \citet{mor09}, \citet{car10},
  \citet{wil10b}, and \citet{mor11}, and were updated when more
  accurate redshifts have become available
  \citep[][]{wil10b,ven12,wil13,wan13}. The green stars illustrate the
  near zone radii of the PS1 $z>6.5$ presented here. The black line
  and shaded region shows the weighted linear fit of
  $R_\mathrm{NZ,corrected}=(7.2\pm0.2)-(6.1\pm0.7)\times(z-6)$. The
  dashes line show the weighted fit derived by \citet{car10} for
  quasars at $z\lesssim6.4$.}
\label{nearzone}
\end{figure*}

Quasar near zones are the regions surrounding distant quasars where the UV
radiation of the central source has ionized the \ion{H}{1}. The size of the
near zone ($R_\mathrm{NZ}$) is a function of, among others, the quasar age,
the flux of ionizing photons and the fraction of neutral H ($f_\mathrm{HI}$)
in the IGM. A study of near zones of $5.7<z<6.4$ quasars was performed by
\citet{car10}, measuring $R_\mathrm{NZ}$ around 27 quasars. To compare the
near zone sizes as function of redshifts, \citet{car10} scaled the measured
$R_\mathrm{NZ}$ to an absolute magnitude of $M_{1450}=-27$:
$R_\mathrm{NZ,corrected}=R_\mathrm{NZ}\times10^{0.4(27+M1450)/3}$. They found
that the ionized region around quasars decreases with increasing redshift
and follows the relation
$R_\mathrm{NZ,corrected}=(7.4\pm0.3)-(8.0\pm1.1)\times(z-6)$. This signals an
increase in $f_\mathrm{HI}$ close to quasars at higher redshifts, although it
is not straightforward to translate a change in $R_\mathrm{NZ}$ to a change in
$f_\mathrm{HI}$ \citep[e.g.,][]{bol07}.

We measured the near zones following the method described in \citet{fan06b}
and also employed by \citet{car10}. The results are shown in Fig.\
\ref{nearzone}a. We derive near zone radii of 1.5, 3.1, and 5.2\,Mpc (proper)
for P167--13, P036+03, and P338+29, respectively. The uncertainty in the
computed near zone (including the uncertainty in the quasar's systemic
redshift derived from \mgii) is about 0.7\,Mpc \citep{car10}. The sizes scaled
to $M_{1450}=-27$ are $R_\mathrm{NZ,corrected}=2.3$, 2.8, and
6.9\,Mpc, respectively.

In Fig.\ \ref{nearzone}b we compare the near zones of the PS1 quasars
with those of $5.7<z<7.1$ quasars from the literature. The PS1 quasars
roughly follow the trend of smaller near zones at higher redshifts. A
weighted linear fit results in a relation
$R_\mathrm{NZ,corrected}=(7.2\pm0.2)-(6.1\pm0.7)\times(z-6)$\,Mpc. Interpreting
the decrease in $R_\mathrm{NZ}$ as an increase in $f_\mathrm{HI}$,
$R_\mathrm{NZ}\propto(1+z)^{-1}f_\mathrm{HI}^{-1/3}$ (e.g., Fan
et~al.\ 2006, but see Bolton \& Haehnelt 2007), the decrease in
$R_\mathrm{NZ,corrected}$ by a factor 6.5 between $z=6$ and $z=7$
implies an increase in the neutral fraction of a factor
$\sim$180. Combined with a measured
$f_\mathrm{HI}\approx2\times10^{-4}$ at $z\sim6$
\citep[e.g.,][]{fan06b}, this suggests $f_\mathrm{HI}\approx0.04$ at
$z=7$, confirming the rapid evolution of $f_\mathrm{HI}$ at $z>6$
\citep[e.g.,][]{fan06b,bol11}. The large spread (a factor $\sim$3) in
(corrected) near zone sizes between individual quasars indicates a
wide range in quasar ages and/or a large variation in $f_\mathrm{HI}$
along different lines of sight.

\section{SUMMARY}
\label{discussion}

\begin{figure}
\includegraphics[width=\columnwidth]{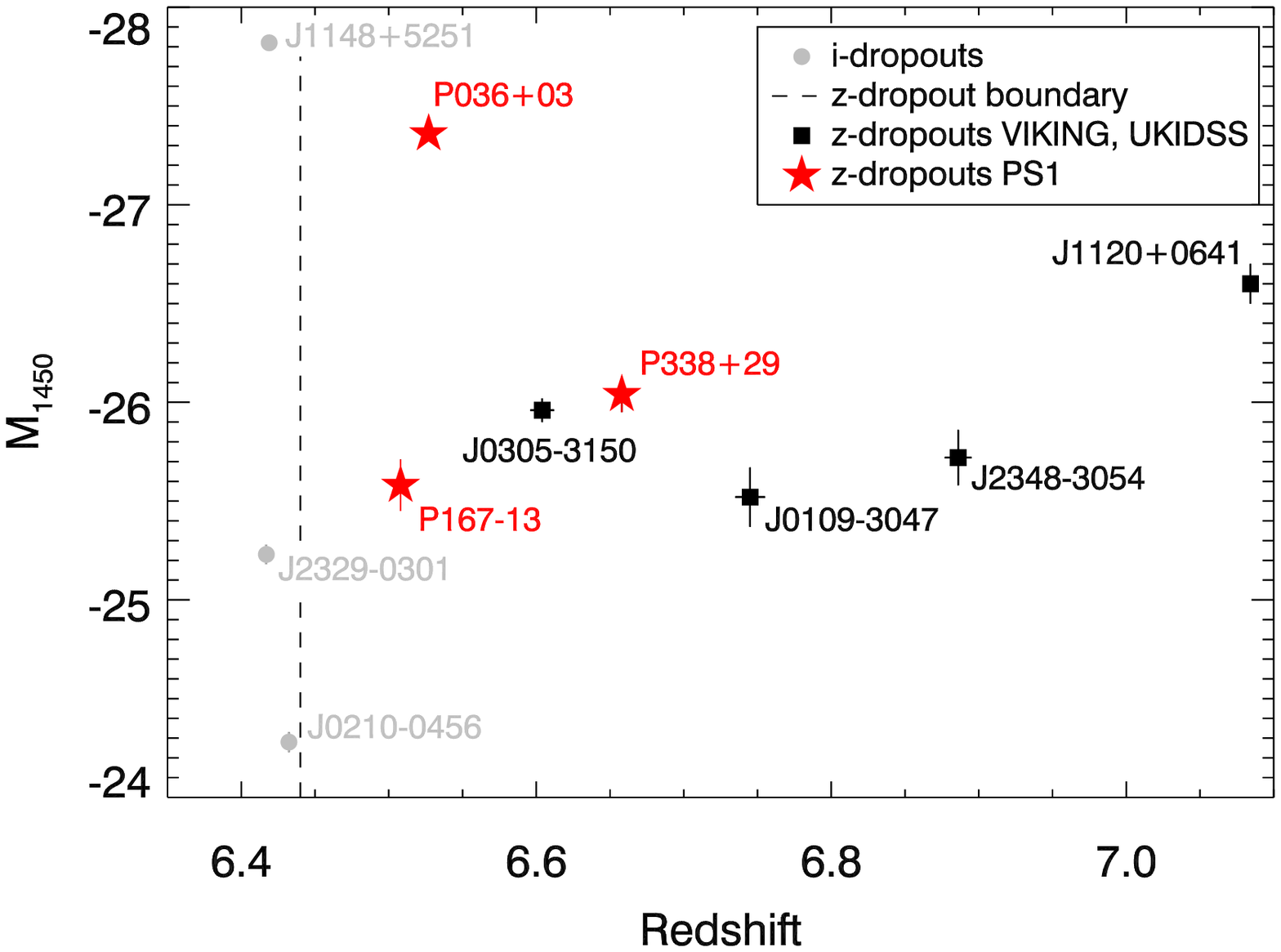}
\caption{Absolute UV magnitude ($M_\mathrm{1450}$) as function of
  redshift for all known quasars at $z>6.4$. The dashed line indicates
  the minimum redshift probed by the $z$-dropout technique
  \citep{ven13}. The black squares are previously known $z$-dropouts
  from \citet{mor11} and \citet{ven13}. The grey circles represent the
  highest redshift quasars discovered in optical surveys
  \citep{fan03,wil07,wil10b}. The new $z>6.5$ quasars presented in
  this work are indicated with red stars.}
\label{mabs}
\end{figure}

We identified three new quasars at redshifts $6.5<z<6.7$ in PS1,
nearly doubling the number of known $z>6.5$ quasars from 4 to 7. The
newly discovered quasars have a wide range of properties (Table
\ref{qsocolors}). The rest-frame UV luminosities are between
$M_{1450}=-25.6$ and $M_{1450}=-27.4$. The brightest of the PS1
quasars is the most luminous quasar discovered at $z>6.5$ so far
(Fig.\ \ref{mabs}), with a luminosity at 1450\,\AA\ close to that of
the bright SDSS quasar J1148+5251 at $z=6.42$ \citep{fan03}. The
faintest PS1 quasar is only marginally brighter than the faintest
$z>6.5$ quasar found in the VIKING survey
\citep[J0109--3047;][]{ven13}. Since the areal coverage of PS1 is more than
10$\times$ larger than that of VIKING, this is very promising for our
continuing PS1 $z$-dropout search.

The PS1 quasars are powered by black holes with estimated masses
between $(0.5-4)\times10^9$\,\msun, based on the \ion{Mg}{2} line
widths and the quasar luminosities. The black holes are accreting in
the range 0.13--1.2 times the Eddington limit. Black hole masses,
accretion rates and (when estimated) \ion{Mg}{2}/\ion{Fe}{2} ratio are
similar to those derived for other $z>6$ quasars
\citep[e.g.,][]{wil10b,der14}.

We derived the ionized region around the quasars and found (luminosity
corrected) near zones between 2.3 and 6.9\,Mpc, in line with the sizes
measured around $5.7<z<6.4$ quasars. By comparing the near zone radii
of quasars between $5.7<z<7.1$, we derive that the average size of the
quasar ionization region decreases by a factor $\sim$6.5 between $z=6$
and $z=7$. This implies a neutral Hydrogen fraction in the IGM of a
few percent at $z=7$, although the scatter in $R_\mathrm{NZ}$ at all
redshifts (a factor 3 between the new quasars) suggest large
variations in $f_\mathrm{HI}$ along different lines of sight.

\acknowledgments 

We thank the referee for carefully reading the manuscript and proving
constructive comments and suggestions.

B.P.V., E.P.F., and F.W. acknowledge funding through ERC grant
``Cosmic Dawn''. E.B. thanks the IMPRS for Astronomy \& Cosmic Physics
at the University of Heidelberg. X.F. and I.D.M. acknowledge support
from US NSF grant AST 11-07682, and R.S. and D.M. from US NSF
grant AST-1109915.

The Pan-STARRS1 Surveys have been made possible through contributions
of the Institute for Astronomy, University of Hawaii, the
Pan-STARRS Project Office, the Max-Planck Society and its
participating institutes, Max-Planck-Institute for Astronomy,
Heidelberg and Max-Planck-Institute for Extraterrestrial Physics,
Garching, The Johns Hopkins University, Durham University, University
of Edinburgh, Queen’s University Belfast, Harvard-Smithsonian Center
for Astrophysics, the Las Cumbres Observatory Global Telescope Network
Incorporated, the National Central University of Taiwan, the Space
Telescope Science Institute, the National Aeronautics and Space
Administration under grant No. NNX08AR22G issued through the Planetary
Science Division of the NASA Science Mission Directorate, the National
Science Foundation under grant AST-1238877, the University of
Maryland, and Eotvos Lorand University (ELTE).

Part of the funding for GROND was granted from the Leibniz-Prize to
Prof. G. Hasinger (DFG grant HA 1850/28-1).

This publication makes use of data products from the {\em Wide-field Infrared
  Survey Explorer}, a joint project of the University of California,
Los Angeles, and the Jet Propulsion Laboratory/California Institute of
Technology, funded by the National Aeronautics and Space Administration.

The LBT is an international collaboration among institutions in the
USA, Italy, and Germany. The partners are: The University of Arizona;
Istituto Nazionale di Astrofisica, Italy; LBT
Beteiligungsgesellschaft, Germany, representing the Max-Planck
Society, the Astrophysical Institute Potsdam, and Heidelberg
University; The Ohio State University; The Research Corporation, on
behalf of The University of Notre Dame, University of Minnesota and
University of Virginia.

{\it Facilities:} 
\facility{PS1 (GPC1)},
\facility{UKIRT (WFCAM)},
\facility{ESO:VISTA (VIRCAM)},
\facility{WISE},
\facility{NTT (EFOSC2, SOFI)},
\facility{Max Planck:2.2m (GROND)},
\facility{CAO:3.5m (OMEGA2000)},
\facility{VLT:Antu (FORS2)}, 
\facility{Magellan:Baade (FIRE)},
\facility{Keck:I (LRIS)},
\facility{MMT (Red Channel Spectrograph)},
\facility{LBT (MODS, LUCI)}.

\end{document}